\documentclass[twocolumn,aps,prb,superscriptaddress]{revtex4}

\usepackage{amsmath,amsthm,amsfonts,graphicx,color,tensor}

\begin{document}
\newcommand{\op}[1]{#1}
\newcommand{\ket}[1]{\left| #1 \right\rangle}
\newcommand{\bra}[1]{\left\langle #1 \right|}
\newcommand{\braket}[2]{\left\langle #1 | #2 \right\rangle}
\newcommand{\ketbra}[2]{\left| #1 \rangle\langle #2 \right|}
\newcommand{\braopket}[3]{\bra{#1}#2\ket{#3}}
\newcommand{\proj}[1]{| #1\rangle\!\langle #1 |}
\newcommand{\expect}[1]{\left\langle#1\right\rangle}
\newcommand{\Entropy}{H}
\newcommand{\KL}[2]{S\left(#1\|#2\right)}
\newcommand{\Tr}{\mathrm{Tr}}
\newcommand{\Rho}{P}
\def\Id{1\!\mathrm{l}}
\newcommand{\cA}{\mathcal{A}}
\newcommand{\cH}{\mathcal{H}}
\newcommand{\cL}{\mathcal{L}}
\newcommand{\cK}{\mathcal{K}}
\newcommand{\cM}{\mathcal{M}}
\newcommand{\cP}{\mathcal{P}}
\newcommand{\cR}{\mathcal{R}}
\newcommand{\cS}{\mathcal{S}}
\newcommand{\cU}{\mathcal{U}}
\newcommand{\reals}{\mathbb{R}}
\newcommand{\grad}{\nabla}
\newcommand{\rhohat}{\hat{\op\rho}}
\newcommand{\rhoMLE}{\rhohat_\mathrm{MLE}}
\newcommand{\rhoHMLE}{\rhohat_\mathrm{H}}
\newcommand{\rhotomo}{\rhohat_\mathrm{tomo}}
\newcommand{\rhotrue}{\rho}
\newcommand{\pvec}[1]{{\bf #1}}
\newcommand{\pMLE}{\pvec{p}_{\mathrm{MLE}}}
\newcommand{\pHMLE}{\pvec{p}_{\mathrm{H}}}
\newcommand{\diff}{\mathrm{d}\!}
\newcommand{\pdiff}[2]{\frac{\partial #1}{\partial #2}}
\def\FCW{1.0\columnwidth}
\def\HCW{0.55\columnwidth}
\def\TPW{0.33\textwidth}

\newcommand{\todo}[1]{\texttt{\color{red}#1}}

\graphicspath{{./}{Figures/}}

\title{Ideal state discrimination with an $O(1)$-qubit quantum computer}

\author{Robin Blume-Kohout}
\affiliation{Theoretical Division, MS-B258, Los Alamos National Laboratory, Los Alamos, NM 87545}
\email{robin@blumekohout.com}
\author{Sarah Croke}
\affiliation{Perimeter Institute for Theoretical Physics, Waterloo, Ontario N2L 2Y5, Canada}
\author{Michael Zwolak}
\affiliation{Department of Physics, Oregon State University, Corvallis, OR 97331}

\maketitle

\textbf{
We show how to optimally discriminate between $K$ distinct quantum states, of which $N$ copies are available, using one-at-a-time interactions with each of the $N$ copies.  While this task (famously) requires joint measurements on all $N$ copies, we show that it can be solved with one-at-a-time ``coherent measurements'' performed by an apparatus with $\log_2 K$ qubits of quantum memory.  We apply the same technique to optimal discrimination between $K$ distinct $N$-particle matrix product states of bond dimension $D$, using a coherent measurement apparatus with $\log_2 K + \log_2 D$ qubits of memory.
}

Quantum state discrimination \cite{CheflesCP00,BarnettAOP09} is the following problem:  Given $N$ quantum systems that were all prepared in one of $K$ distinct states $\ket{\psi_1},\ldots,\ket{\psi_K}$, decide in which state they were prepared.  Finding the optimal measurement is a straightforward convex program, in principle.  But when $N>1$ copies of $\ket{\psi_k}$ are available, it is usually a joint measurement on all $N$ copies.  Such measurements can be prohibitively difficult.  Observing each of the $N$ copies independently yields a strictly lower probability of success \cite{HelstromBook76,PeresPRL91}.  This contrasts starkly with the corresponding classical problem of distinguishing $K$ distinct probability distributions, where one-at-a-time observations are completely sufficient.

In this paper, we demonstrate that by slightly relaxing the usual meaning of ``observation'', it \emph{is} possible to do optimal discrimination using one-at-a-time observations, which restores a pleasing symmetry with the classical case.  A quantum measurement conventionally comprises: (i) a controlled unitary interaction between a system $\cS$ and an apparatus $\cA$; (ii) decoherence on $\cA$, which forces its state into a mixture of ``pointer basis'' states \cite{ZurekRMP03}; and (iii) experimental readout of the classical result from $\cA$ (arguably accompanied by ``collapse'' of $\cA$'s state).  We relax this prescription by making $\cA$ a quantum information processor (QIP) -- basically a very small (perhaps just 1 qubit) non-scalable quantum computer.  We protect $\cA$ from decoherence and avoid reading out any information until the very end of the protocol.  What remains is a \emph{coherent measurement}, a unitary interaction between $\cS$ and $\cA$ that transfers information from $\cS$ to $\cA$.

We begin by showing how to realize optimal discrimination between $N$ copies of $K$ pure states with one-at-a-time coherent measurements, using a $\log_2K$-qubit QIP.  Next, we apply the same technique to optimal discrimination of many-body \emph{matrix product states} (MPS).  Our protocol distinguishes between $K$ distinct $N$-particle MPS with bond dimension $D$, and uses a $(\log_2K+\log_2D)$-qubit QIP.  Finally, we combine our first two results to get a protocol for discriminating between $M$ copies of $K$ distinct MPS using a $\log_2K+\log_2D$ qubit QIP.

\section{Discriminating $N$-copy states}
\label{sec:ncopy} 

Suppose we are given $N$ quantum systems ($\cS_1\ldots \cS_N$) with $d$-dimensional Hilbert spaces $\cH_n$, and a promise that they were all identically prepared in one of $K$ nonorthogonal states $\{\ket{\psi_1}\ldots\ket{\psi_K}\}$.  Their joint state is
$\ket{\psi_k}^{\otimes N}\in\cH^{\otimes N}$, with $k$ unknown.  Identifying $k$ with maximum success probability requires a joint measurement on all $N$ samples.  Non-adaptive one-at-a-time measurement cannot achieve the optimal success probability.  For $K=2$ candidate states, there is an adaptive local measurement scheme that achieves the optimal success probability \cite{AcinPRA05}, but no such protocol has been found for $K>2$ (despite some effort \cite{WoottersPrivate} -- which suggests, but certainly does not prove, that no such protocol exists).

All the information about $k$ is contained in a $K$-dimensional subspace 
\begin{equation}
\cK_N = \mathrm{Span}\left(\left\{\ket{\psi_k}^{\otimes N}\right\}\right). \label{eq:KN}
\end{equation}
So while the optimal measurement is a joint measurement, it does not need to explore the majority of $\cH$.  We will implement it by rotating the entire subspace $\cK$ into the state space of our $K$-dimensional QIP $\cA$ (the coherent measurement apparatus).  We do so via sequential independent interactions between $\cA$ and each of the $N$ samples $\cS_n$, ``rolling up'' all information about $k$ into $\cA$.

$\cA$ is initially prepared in the $\ket{0}$ state.  We bring it into contact with $\cS_1$, and execute a SWAP gate between $\cS_1$ and the $\{\ket{0},\ldots,\ket{d-1}\}$ subspace of $\cA$.  This transfers all information from the first sample into $\cA$, leaving $\cS_1$ in the $\ket{0}$ state.

Now we bring $\cA$ into contact with $\cS_2$.  Their joint state is $\ket{\psi_k}^{\otimes 2}$, although we do not know $k$.  But we do know that the state lies within $\cK_2 = \mathrm{Span}(\{\ket{\psi_k}^{\otimes 2}\})$, (see Eq. \ref{eq:KN}),
whose dimension is at most $K$.  A basis $\{\ket{\phi_j}:\ j=1\ldots K\}$ for this space can be obtained by Gram-Schmidt orthogonalization.  We apply a unitary interaction\footnote{We have only defined $U_2$ on the subspace of interest.  For completeness, it can be extended to the complement, $\cH_A\otimes\cH_2 / \cK_2$, in any convenient manner.} between $\cA$ and $\cS_2$,
\begin{equation}
U_2 = \sum_j{\ketbra{0_{\cS_2}j_{\cA}}{\phi_j}}. \label{eq:U2}
\end{equation}
It rotates $\cK_2$ into $\{\ket{0}\}_{\cS_2}\otimes\cH_{\cA}$, which places all the information about $k$ in $\cA$ and decouples $\cS_2$.  ($\cS_2$ is left with \emph{no} information about $k$ if and only if $\cA$ is left with \emph{all} the information about $k$.)  $\cA$ is now in one of $K$ possible states $\ket{\psi_k^{(2)}}$, which (as a set) are unitarily equivalent to $\{\ket{\psi_k}^{\otimes 2}\}$ -- e.g., $\braket{\psi_j^{(2)}}{\psi_k^{(2)}} = \braket{\psi_j}{\psi_k}^2$.

The rest of the algorithm is now fairly obvious; we move on and interact $\cA$ with $\cS_3$ in the same way, etc, etc.  At each step, when $\cA$ comes into contact with $\cS_n$, their joint state is $\ket{\psi_k^{(n-1)}}\otimes\ket{\psi_k}$.  These $K$ alternatives span a $K$-dimensional space $\cK_n$ (see Eq. \ref{eq:KN}), spanned by a basis $\left\{\ket{\phi^{(n)}_j}\right\}$, which is then rotated into $\{\ket{0}\}_{\cS_n}\otimes\cH_{\cA}$ by applying
$$U_n = \sum_j{\ketbra{0_{\cS_n}j_{\cA}}{\phi^{(n)}_j}}.$$
Each sample system is left in the $\ket{0}$ state, indicating that all its information has been extracted.  After every sample has been sucked dry, we simply measure $\cA$ to extract $k$.  This final measurement can be efficiently computed via convex programming, since $\cA$ is only $K$-dimensional.

The sequence of coherent measurement interactions is independent of what sort of discrimination we want to do -- e.g., minimum-error \cite{HelstromBook76}, unambiguous discrimination \cite{IvanovicPLA87,DieksPLA88,PeresPLA88,JaegerPLA95}, maximum-confidence \cite{CrokePRL06}, etc -- because $\cK_N$ is a \emph{sufficient statistic} for any inference about $k$, and our protocol simply extracts it whole, leaving the decision rule up to the final measurement on $\cA$.  As in the classical case, data gathering can now be separated from data analysis.

This protocol can be modified to discriminate non-symmetric product states -- e.g., $\ket{\psi_1}\otimes\ket{\psi_2}\otimes\ldots\ket{\psi_N}$ vs. $\ket{\phi_1}\otimes\ket{\phi_2}\otimes\ldots\ket{\phi_N}$.


\section{Matrix product state discrimination}
\label{sec:MPS}
                    
The information about $k$ can be ``rolled up'' using sequential interactions because it is contained in a subspace $\cK_N$ with Schmidt rank\footnote{By this we mean that, given any pure or mixed state $\rho$ on $\cK_N$, if we trace out $\{\cS_1\ldots\cS_n\}$, the reduced state for $\{\cS_{n+1}\ldots\cS_{N}\}$ has rank at most $K$.} at most $K$ across any division of the $N$ systems.  Low Schmidt rank is critical. Consider distinguishing between two states that are each maximally entangled between the first $N/2$ samples and the last $N/2$ samples.  They lie in a 2-dimensional subspace, but it is \emph{not} accessible through our protocol.  The first $N/2$ samples are maximally entangled with the rest, so their reduced state has rank $d^{N/2}$.  At least $N\log_2d/2$ qubits would be needed to store the information extracted from the first $N/2$ samples.

But whenever the Schmidt rank condition is satisfied, a variation of our algorithm will work.
For product states (above), each state has Schmidt rank 1, and the span of $K$ such states has Schmidt rank at most $K$.

This property of low Schmidt rank is generalized by \emph{matrix product states} (MPS) \cite{FannesCMP92,VerstraeteAIP08}.  An $N$-particle MPS with \emph{bond dimension} $D$ is guaranteed to have Schmidt rank at most $D$ across any division of the 1D lattice.  Thus, the span of $K$ such MPS, each with bond dimension $\leq D$, has Schmidt rank at most $DK$.  We denote such a subspace,
$$\cK = \mathrm{Span}(\{\ket{\psi_k}\}),$$
a \emph{matrix product subspace} with bond dimension $DK$.  Such a set of MPS can be distinguished optimally with coherent measurements and $(\log_2D+\log_2K)$ qubits.

\begin{figure}[t]
\begin{centering}
\includegraphics[width=8cm]{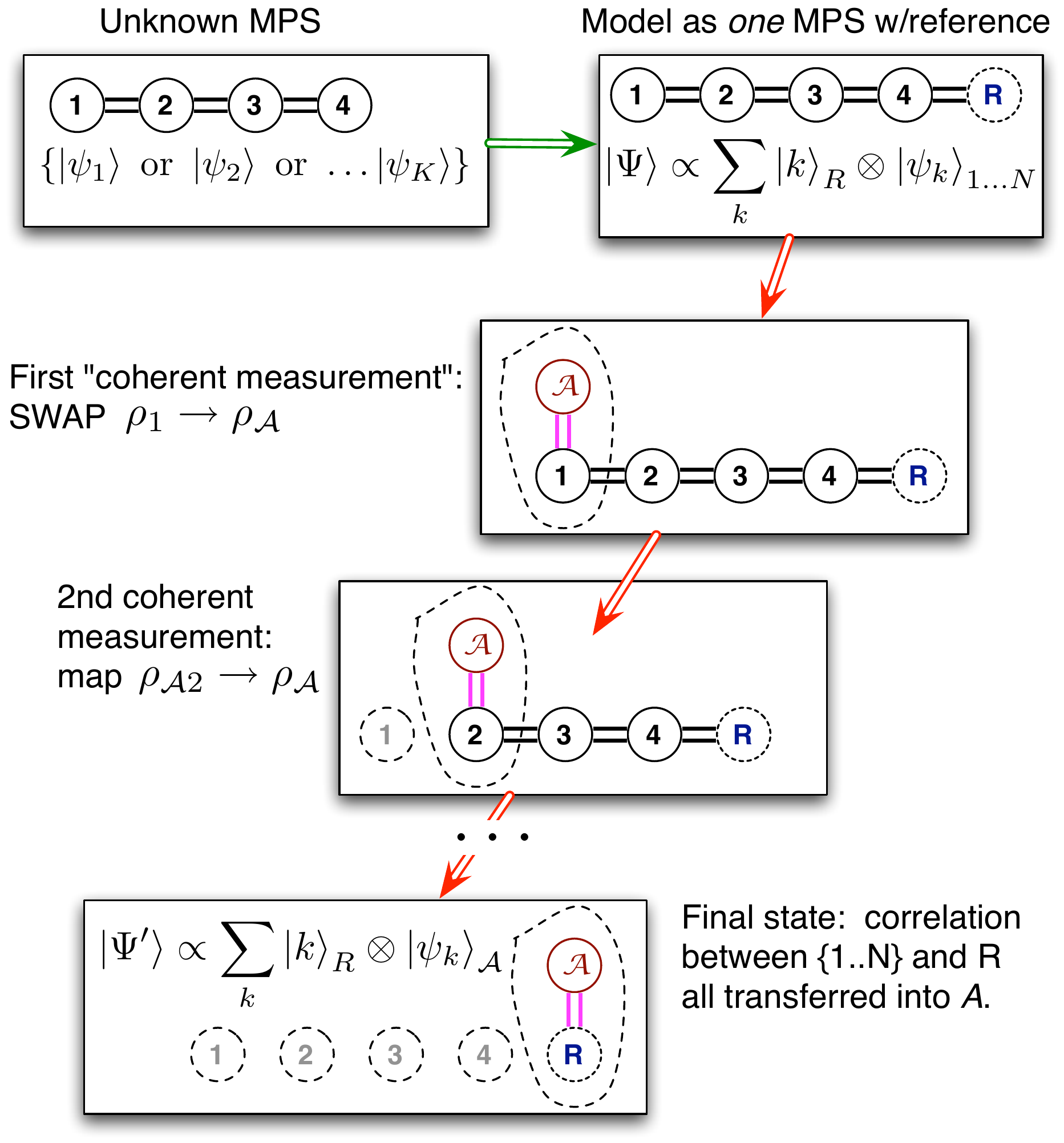}
\par\end{centering}
\caption{Our protocol represents an unknown MPS $\ket{\psi_k}$ from a set $\{\ket{\psi_1}\ldots\ket{\psi_K}\}$ by its \emph{purification} -- a single MPS $\ket{\Psi}$ involving a fictitious reference $R$.  The algorithm then successively decorrelates each sample $\cS_n$ from the rest, storing $\cS_n$'s correlations with the remainder of the lattice in the ``apparatus'' $\cA$.  Ultimately, all information about $k$ (i.e., correlation with $R$) is contained in $\cA$, which can be measured.\label{fig:mps}}
\end{figure}

Our algorithm is a straightforward generalization of the one for product states, and proceeds as shown in Figure~\ref{fig:mps}.  First, we represent the MP subspace $\cK$ by its \emph{purification} -- a single MPS $\ket{\Psi}$ for $\cS_1\ldots \cS_N$ \emph{and} a fictitious reference system $R$,
\begin{equation}
\ket{\Psi}\propto \sum_k \ket{k}_R\ket{\psi_k}_{\cS_1\ldots \cS_N}, \label{eq:MPsub}
\end{equation}
Information about $k$, which is contained in $\cK$, equates to correlation with the imaginary $R$.

Now, we initialize $\cA$ in the $\ket{0}$ state, then SWAP its state with that of $\cS_1$ (the first lattice site).  This decouples $\cS_1$, leaving $\cA\otimes \cS_2\ldots\cS_N \otimes R$ in a matrix product state,
\begin{equation}
\ket{0}_\cA\ket{\Psi}_{\cS_1\ldots \cS_N, R} \to \ket{0}_1\ket{\Psi}_{\cA,\cS_2\ldots \cS_N, R},
\end{equation}
with Schmidt rank no greater than $DK$.


Now, to roll up each successive site $\cS_n$ ($n=2, \ldots, N$), we find the Schmidt decomposition of the current state between $\cA\otimes\cS_n$ and the remainder of the lattice, write it (generically) as
\begin{equation}
\sum_{j=1}^{DK} c_j {\ket{\mu_j}_{\cA\cS_n}\ket{\nu_j}_{\cS_{n+1}\ldots\cS_N,R}},
\end{equation}
and apply a unitary operation\footnote{As in Eq. \ref{eq:U2}, this unitary is specified only on the subspace of interest, and can be completed in any convenient fashion.} to $\cA\otimes\cS_n$,
\begin{equation}
U_n = \sum_j{\ket{0_{\cS_n} j_\cA}\bra{\mu_j}},
\end{equation}
which decouples $\cS_n$ and leaves all the information previously in $\cS_n\otimes\cA$ in $\cA$.  Doing this successively at each site decorrelates all the $\cS_n$, and we are left in the state 
$$\ket{\Psi'} \propto \sum_k{\ket{k}_R\ket{\psi_k}_{\cA}},$$
with all information about $k$ in $\cA$, where it can be extracted by a simple measurement.


Recent proposals for local tomography \cite{CramerNC10} are also based on sequential interactions.  Our protocol, with coherent measurements, offers a tremendous efficiency advantage (at the cost of requiring a small QIP!).  It can distinguish near-orthogonal MPS states with a single copy, whereas local tomography requires $O(N)$ copies.  Distinguishing non-orthogonal states requires multiple ($M$) copies.  To apply our algorithm, we simply line up the copies (they do \emph{not} have to exist simultaneously) and treat them as a single $NM$-particle MPS of bond dimension $D$.

\section{Mixed state discrimination and tomography}

In the context of $N$-copy states (Section \ref{sec:ncopy}), one may ask:
\begin{enumerate}
\item Can coherent measurement be used to discriminate \emph{mixed} states, i.e. $\rho_k^{\otimes N}$?
\item Can coherent measurement be used for full state tomography (rather than discrimination)?
\end{enumerate}
The answer to both is ``Yes, but it seems to require an $O(\log N)$-qubit apparatus.''  This is a very favorable scaling, but less remarkable (and less immediately useful) than the $O(1)$ scaling for pure state discrimination.

This is possible because the \emph{order} of the samples is completely irrelevant.  As we scan through the samples, we can discard ordering information, keeping only a sufficient statistic for inference about $\rho$.  The \emph{quantum Schur transform} does this\cite{BaconPRL06}.  It is based on Schur-Weyl duality \cite{HarrowThesis}, which states that because the $N$-copy Hilbert space $\cH_d^{\otimes N}$, permutations of the samples commute with collective rotations of all $N$ samples, the Hilbert space can be refactored as
$$\cH^{\otimes N} = \bigoplus_{\lambda}{\cU_\lambda \otimes \cP_\lambda}.$$
The $\cU_\lambda$ factors are \emph{irreducible representation spaces} (irreps) of $SU(d)$, the $\cP_\lambda$ factors are irreps of $S_N$, and $\lambda$ labels the various irreps.  The Schur transform can be implemented by a unitary circuit that acts sequentially on the samples, mapping $N$ qudit registers into three quantum registers containing (respectively) $\lambda$, $\cU$, and $\cP$:
$$\cH^{\otimes N} \to \cH_{\lambda}\otimes\cH_\cU\otimes\cH_\cP.$$

The ``ordering'' register $\cH_\cP$ accounts for nearly all the Hilbert space dimension of $\cH_d^{\otimes N}$, and since it is irrelevant to inference it can be discarded as rapidly as it is produced.
What remains to be stored in $\cA$ is:
\begin{enumerate}
\item a ``label'' register $\lambda$ (a sufficient statistic for the spectrum of $\rho$),
\item a $SU(d)$ register $\cU$ (a sufficient statistic for the eigenbasis of $\rho$).
\end{enumerate}
The $\lambda$ register requires a Hilbert space with dimension $\geq$ the number of Young diagrams with $N$ boxes in at most $d$ rows, which is approximately
$$\mathrm{dim}(\lambda) \approx \frac{1}{d!}\binom{N}{d}.$$
The $\cU$ register must hold the largest $N$-copy irrep of $SU(d)$, whose size can be calculated using hook-length formulae \cite{Georgi}, and upper bounded by
$$\mathrm{dim}_{\mathrm{max}}(\cU) = (N+d-1)^{\frac12d(d-1)}.$$
Together, these registers require $O(d^2\log N)$ qubits of memory (although for pure state tomography, $O(d\log N)$ qubits of memory are sufficient).

$O(\log N)$ memory appears to be \emph{necessary} for optimal accuracy.  Consider the simplest possible case -- discrimination of two \emph{classical} 1-bit probability distributions
$$\binom{p}{1-p}\mathrm{\ vs.\ }\binom{q}{1-q}.$$
The sufficient statistic is frequencies of ``0'' and ``1'', $\{n,N-n\}$.  For any given problem, there is a threshold value $n_c$ such that the answer depends only on whether $n<n_c$, so only one bit of information is required.  However, extracting that bit via sequential queries requires storing $n$ exactly at every step (using $O(\log N)$ bits of memory).  Any loss of precision could cause a $\pm1$ error at the final step, and thus a wrong decision.  In this example, classical storage is sufficient.  But in the general case, where the candidate $\rho_k$ do not commute, no method is known to compress the intermediate data into classical memory without loss (previous work suggests it is probably impossible \cite{KoashiPRL02}).

\section{Discussion:  applications, implementations, and implications}

Quantum information science is rife with gaps between what is theoretically achievable and what is practically achievable.  Our algorithm eliminates performance gaps for pure state discrimination with local measurements -- but it requires a new kind of measurement apparatus with at least 1 controllable qubit of quantum storage.  Its utility depends on its applications, and on the difficulty of implementation.

\vspace{0.1in}\noindent\textbf{APPLICATIONS:}  One \emph{immediate} application of our protocol is detection of weak forces and transient effects.  A simple force detector (e.g., for magnetic fields) might comprise a large array of identical systems (e.g., $\ket{\downarrow}$ spins).  Each system is only weakly perturbed by the force, so information about the force is distributed across the entire array.  Our algorithm efficiently gathers up that information with no loss -- whereas local measurements with classical processing waste much information.

A more sensitive $N$-particle ``antenna'' would incorporate entanglement between the $N$ particles \cite{BartlettRMP07}.  High sensitivity can be achieved by simple MPS states with $D=2$, like N00N states \cite{DowlingCP08},
$$\ket\psi = \frac{\ket{N,0} + \ket{0,N}}{\sqrt2} = \frac{\ket{\uparrow}^{\otimes N}+\ket{\downarrow}^{\otimes N}}{\sqrt2}.$$
Collective forces do not change $D$, so the final states to be discriminated are also MPS.  Our approach can discriminate such states \emph{and} it can be used to prepare them, by running the ``rolling up'' process in reverse \cite{SchonPRL05}.

More ambitious applications include direct probing of many-body states, to test a particular MPS ansatz for a lattice system, or to characterize results of quantum simulations in optical lattices or ion traps.  Without fully scalable quantum computers that can couple directly to many-body systems, coherent measurements may be the only way to efficiently probe complex $N$-particle states.  Our protocol does not obviously scale to \emph{PEPS}, the higher-dimensional analogues of MPS \cite{VerstraeteAIP08}.  Like MPS, PEPS obey an \emph{area law} -- entanglement across a cut scales not with the volume of the lattice ($N$), but with the area of the cut.  For a 1-dimensional MPS on $N$ systems, any cut has area 1, so the Schmidt rank scales as $O(1)$, and our algorithm requires an $O(1)$ qubit QIP.  Rolling up a general PEPS on an $n$-dimensional lattice would require $O(N^{\frac{n-1}{n}})$ bits of quantum memory.  However, some PEPS can be sequentially generated \cite{BanulsPRA08}, and are likely amenable to our protocol.

\vspace{0.1in}\noindent\textbf{IMPLEMENTATIONS:}  Engineering requirements for a coherent measurement apparatus are achievable with near-future technology.  $\cA$ should be a clean $K$-dimensional quantum system with:
\begin{enumerate}
\item Universal local control,
\item Long coherence time relative to the gate timescale,
\item Controllable interaction with an external $d$-dimensional ``sample'' system,
\item Sequential coupling to each of $N$ samples,
\item Strong measurements (which may be destructive).
\end{enumerate}
$K=2$ is sufficient for proof-of-principle, but $K\geq3$ would be more exciting because adaptive local measurements can discriminate $K=2$ states.

These requirements are much weaker than those for scalable quantum computing.  Coherent measurement could be an early application for embryonic quantum architectures.  Furthermore, scalability is not required, just a single $K$-dimensional system.  Only local control has to be universal, since the interaction with external systems is limited.  Error correction is not mandatory, for coherence need only persist long enough to interact with each of the $N$ systems of interest.  Since measurements are postponed until the end, they can be destructive.

We do require $\cA$ to be portable -- i.e., sequentially coupled to each of the $N$ samples -- whereas a quantum computer can be built using only nearest neighbor interactions.  Fortunately, most proposed architectures have selective coupling either through frequency space (NMR, ion traps with a phonon bus) or physical motion of the qubits (some ion traps) or flying qubits (photonic architectures).  Devices that are not viable for full-scale quantum computing may be even better for coherent measurement.  For example, an STM might pick up and transport a single coherent atomic spin along an array of sample atoms, interacting sequentially with each of them.

\vspace{0.1in}\noindent\textbf{IMPLICATIONS:}  Coherent measurements are a genuinely new way to gather information.  We have not just removed collapse from standard quantum measurements!  That kind of coherent measurement is used already in quantum error correction, where it's common to replace a measurement of $X$ with a controlled unitary of the form
\begin{equation}
U_{\cS\cA} = \sum_x{\proj{x}_\cS\otimes U^{(x)}_\cA}. \label{eq:CU}
\end{equation}
Such unitaries transfer information about a specific observable $X$ from $\cS$ to $\cA$. For appropriate $\ket{\psi_0}_\cA$ and $U^{(x)}_\cA$, later measurements of $\cA$ produce exactly the same result as if $\cS$ had been measured directly.  The coherent measurements in our discrimination protocols are \emph{not} of this form.  They do not measure (i.e., transfer information to $\cA$ about) a specific basis.  For example, in $N$-copy state discrimination, $\cA$ interacts with the first sample by a SWAP operation, which has no preferred basis.  Later interactions are also not of controlled-U form (Eq. \ref{eq:CU}).

One might ask where the ``measurement'' occurs, since the interaction between $\cS$ and $\cA$ is purely unitary.  The essence of measurement is that an observer or apparatus gains information.  Quantum measurements are usually construed as mysterious processes that consume quantum states and excrete specific, definite measurement outcomes.  Quantum theories of measurements usually represent them as (i) unitary interaction, (ii) decoherence and superselection, and finally (iii) wavefunction collapse or splitting of the universe \footnote{In the many worlds interpretation of quantum mechanics, each of the possible outcomes occurs in a subset of the possible universes; this does not seem to conclusively explain why we observe \emph{our} particular outcome in \emph{our} universe.}.  Our results suggest that unitary interaction (the only part of this sequence that is really understood) can stand alone as an information-gathering ``measurement.'' And by avoiding decoherence, we can gather information strictly \emph{more} effectively.

Decoherence is ubiquitous in human experience.  But in its absence, there is no compelling reason why gathering information must be accompanied by collapse or definite outcomes.  The whole point of quantum information science is to produce devices that do \emph{not} decohere, and that can \emph{process} information more efficiently than classical computers.  The central message of this paper is that they can also \emph{gather} information more efficiently.  Unfettered by decoherence, they may still be constrained by locality.  For such devices, coherent measurements are the natural way to gather information.

\vspace{0.1in}\noindent\textbf{ACKNOWLEDGEMENTS}:  We are grateful for comments by semi-anonymous QIP 2012 referees.  This work was supported by the US Department of Energy through the LANL/LDRD program (RBK and MPZ), as well as by Perimeter Institute for Theoretical Physics (RBK and SC). Research at Perimeter Institute is supported by the Government of Canada through Industry Canada and 
by the Province of Ontario through the Ministry of Research \& Innovation.

\end{document}